\documentclass[pdflatex,sn-mathphys-num]{sn-jnl}% Math and Physical Sciences Numbered Reference Style
\usepackage{graphicx}
\usepackage{tabulary}
\usepackage{multirow}%
\usepackage{amsmath,amssymb,amsfonts}%
\usepackage{amsthm}%
\usepackage{mathrsfs}%
\usepackage[title]{appendix}%
\usepackage{xcolor}%
\usepackage{textcomp}%
\usepackage{manyfoot}%
\usepackage{booktabs}%
\usepackage{algorithm}%
\usepackage{algorithmicx}%
\usepackage{algpseudocode}%
\usepackage{listings}%
\usepackage{array}
\usepackage{subcaption}
\begin{document}

\title[Observationanal and Thermodynamic Analysis of Power Law form of $f (Q)$ gravity]{Observationanal and Thermodynamic Analysis of Power Law form of $f (Q)$ gravity}


\author[1]{\fnm{Agnidipto} \sur{Bhattacharya}}\email{bhattacharyaagnidipto@gmail.com}
\author[2]{\fnm{Shinjini} \sur{Ghosh}}\email{ghoshshin@gmail.com}
\author*[3]{\fnm{Atreyee} \sur{Biswas}}\email{atreyee11@gmail.com}
\author[4]{\fnm{Sujay} \sur{Kr. Biswas}}\email{sujaymathju@gmail.com}



\affil*[1,3]{\orgdiv{Department of Applied Mathematics, Maulana Abul Kalam Azad University of Technology, W.B} }
\affil[2]{\orgdiv{Department of Mathematics and Statistics, Heramba Chandra College, , West Bengal, India}}
\affil[4]{\orgdiv{Department of Mathematics}, \orgname{University of North Bengal, W.B}}


\abstract
{
In this study, the power law form of f(Q) gravity $(f(Q)=Q+mQ^n)$ has been studied in view of statistical analysis with current observed data and irreversible thermodynamics. The best fit values of  model specific parameters are found with the use of of 31point Hubble data, Patntheon type Supernva and 1721 point SH0ES data and a comparative study has been performed between $f(Q)$ and standard $\Lambda$CDM model. The thermodynamic consequence using standard Eckart theory of non-equilibrium thermodynamics has been examined independently for general form of $f(Q)$ gravity followed by power law form of $f(Q)$ where the horizon entropy is calculated using unified law of thermodynamics. Suitable range of power law model specific parameters has been explored through investigating validity of generalized second law of thermodynamics and thermodynamic equilibrium. Finally evolution nature of cosmographic parameters are studied within the specific range of $m,n$ which is compatible with the results obtained from both statistical and thermodynamical analysis.
}




\maketitle
\section{Introduction}
From several independent cosmological probes like Type Ia supernovae (SNeIa) \cite{Riess 1998}, the cosmic microwave background (CMB) \cite{Bennet 2003}, and large-scale structure (LSS) surveys \cite{Hawkins 2003} etc points to the fact that the universe is currently undergoing a phase of accelerated expansion. This late time acceleration of universe is dealt mainly through two approaches. In one approach the existence of an exotic matter called dark energy having negative pressure is considered while in the other approach Einstein's general relativity theory is modified giving rise to theoty of modified gravity. There are various ways in modifying Einstein's theory of general relativity. One way consists of extending the Einstein-Hilbert Lagrangian with additional terms while preserving its underlying geometric (Riemannian) framework. This gives rise to various classes of modified gravities such as $f(R)$ gravity\cite{Soritiu 2010}, Gauss-Bonnet and $f(G) $ gravity\cite{Lokhare 2024, Garcia 2011}, Weyl gravity\cite{Wheeler 2014}, etc. A separate class of modified gravity theories emerges from the torsion-based formulation, known as modified teleparallel gravity. This includes $f(T)$ gravity \cite{Cai 2016, Palianthanasis 2016}, $f(T,T_G)$ gravity \cite{Kofinas 2014}, and scalar-torsion theories \cite{Hohmann 2018, Hohmann1 2018}, among others. Yet another formulation relies on the non-metricity $Q$ of the connection, in which curvature and torsion are both set to zero while only non-metricity remains non-vanishing. This allows gravity to be described geometrically through non-metricity alone, giving rise to what is known as the Symmetric Teleparallel Equivalent of General Relativity (STEGR), or $f(Q)$ gravity. $f(Q)$ gravity was first introduced by Jim´enez, Heisenberg, and Koivisto \cite{Jimenez 2020}. Afterwards during last few years f(Q) gravity and and its generalizations have attracted much interest to researchers. Lazkoz et al. \cite{Lazkoz 2019} derived set of constraints on $f(Q)$ gravity using data from recent observations while S. Mondal et al \cite{Mandal 2020} used energy conditions to impose restrictions on model parameters of $f(Q)$ models. One can go through the literature \cite{Shabani 2024} for more reviews on $f(Q)$ gravity. \\
In this work, we have studied one particular $f(Q)$ model viz., the power law $f(Q)$ model in light of  theory of thermodynamics, observational and cosmographic analysis. This model can explain the accelerated expansion of the universe at late times while remaining consistent with constraints from Big Bang Nucleosynthesis (BBN) \cite{Anagnostopoulos 2023}. . In \cite{Shabani 2024}  S. Mondal et al imposed limit on parameters of power law model using energy conditions while P. K. Sahoo et al \cite{Sahoo 2023} discussed this particular form of $f(Q)$ gravity constraining the model by constant sound speed in view of current observed data, thermodynamics and cosmographic analysis. In this work our objective is to investigate the power law form of $f(Q)$ gravity without limiting the model by any prior constraint. The model is examined primarily in respect of of observational analysis and thermodynamics. Our objective is to establish correspondence between the results obtained from these two analysis. However, it is to be mentioned that we have considered irreversible thermodynamics in our present work in place of traditional reversible thermodynamics.\\  
A remarkable discovery in the 1970s by Stephen Hawking \cite{Hawking1975} revealed that black hole behaves as genuine thermodynamic objects, emitting thermal radiation known as Hawking radiation. This finding proved to be a turning point in cosmological research. Thereafter Bardeen et al. (1973) \cite{Bardeen 1973} established a correspondence between the four laws of thermodynamics and the four laws of black hole mechanics, wherein temperature and entropy correspond, respectively, to the surface gravity and horizon area of the black hole. Subsequently, Jacobson (1995) \cite{Jacobson 1995} proved that Einstein's field equations could be derived directly from the first law of thermodynamics, while Padmanabhan (2002) \cite{Padmanabhan 2002} showed the converse;  he for a general static, spherically symmetric spacetime derived the first law of thermodynamics starting from Einstein's field equations. Together, these two results suggest that the thermodynamic description is not confined to black holes alone but extends to general spacetime  — implying that the universe as a whole may likewise be treated as a thermodynamical system. However, universal thermodynamics must be treated as an irreversible process — rather than a reversible or quasi-reversible one. This was first recognized by Eling et al. \cite{Eling 2006} in their attempt to derive Einstein's field equations from the first law of thermodynamics within the framework of f(R) gravity. They showed that a non-equilibrium thermodynamic approach is essential for this successful deduction . Furthermore, a non-equilibrium thermodynamic formalism is indispensable for explaining several key phenomena during the evolution of universe, such as the neutrino decoupling in the early universe, photon-matter decoupling during recombination era and nucleosynthesis.
 We organize this paper in the following way:\\
 Section 2 deals with the general prescription of $f (Q)$ gravity theory has been illustrated while in section 3, $f(Q)$ gravity model is examined through statistical analysis. In particular, best fit values of model parameters and their influence on evolution dynamics of universe has been studied using current observed data. Section 4 deals with the study of $f(Q)$ gravity model in context of  irreversible thermodynamics. In particular, here general theory of irreversible thermodynamics is discussed and Unified law of thermodynamics has been applied to find horizon entropy in $f(Q)$ gravity framework. Then for a general $f (Q)$ gravity model, we examined the validity of the generalized second law of thermodynamics (GSLT) and Thermodynamic equilibrium (TE) and then applied the obtained results on our specific $f(Q)$ model(Power law) to find acceptable range of model parameters.Section 5 studies the universe's evolution dynamics via cosmographic parameters, restricted to the parameter ranges obtained from the observational and thermodynamic analyses. Lastly  section 6 is dedicated for summary and conclusion.

\section{ f(Q) gravity framework}

In $f(Q)$ gravity the action is given by
\begin{equation}
S=\int [-\frac{1}{16\pi G f(Q)}+\mathcal{L}_m]\sqrt{-g}d^4x \label{action}
\end{equation}
where $\mathcal{L}_m$ is the matter Lagrangian density and  $g$ being the determinant of the metric $g_{\mu\nu}$. $f(Q)$ is an arbitrary function of the non-metricity scalar $Q$ where\\
\begin{equation}
\label{kk2}
Q=-\frac{1}{4}Q_{\alpha\beta\gamma}Q^{\alpha\beta\gamma}+\frac{1}{2}Q_{\alpha\beta\gamma}Q^{\gamma\beta\alpha}+\frac{1}{4}Q_{\alpha}Q^{\alpha}-\frac{1}{2}Q_{\alpha}\bar{Q}^{\alpha}
\end{equation}
Here $Q_\alpha=Q_\alpha^{\mu_\mu}$ and $\bar{Q}^\alpha=Q_\mu^{\mu\alpha}$ are obtained from
contractions of the non-metricity tensor $Q_{\alpha\mu\gamma}=\nabla_\alpha g_{\mu\nu}$\\
Putting $8\pi G = 1$ for simplicity from calculational point of view and varying the action (\ref{action}) leads to the following field equations-
\begin{equation}
    \frac{2}{\sqrt{-g}}\nabla_\alpha\{\sqrt{-g}g_{\beta\nu}[-\frac{1}{2}L^{\alpha\mu\beta}+\frac{1}{4}g^{\mu\beta}(Q^\alpha-\bar{Q}^\alpha)-\frac{1}{8}(g^{\alpha\mu}Q^\beta+g^{\alpha\beta}Q^\mu]\}\nonumber
\end{equation}
\begin{equation}
        +f_Q[-\frac{1}{2}L^{\mu\alpha\beta}-\frac{1}{8}(g^{\mu\alpha}Q^\beta+g^{\mu\beta}Q^\alpha)+\frac{1}{4}g^{\alpha\beta}(Q^\mu-\bar{Q}^\mu)]Q_{\nu\alpha\beta}+\frac{1}{2}\delta^\mu_\nu f=T^\mu_{\nu}\label{Einstein Field eq}
\end{equation}
where, $L^\alpha_{\mu\nu}=\frac{1}{2}Q^\alpha_{\mu\nu}-Q^\alpha_{(\mu \nu)}$
is the deformation tensor and $T_{\mu\nu}$ represents the energy-momentum tensor. $f_Q=\frac{\partial f}{\partial Q}$.\\
However, we assume our universe to be flat, homogeneous and isotropic described by the FLRW metric :
\begin{equation}
    ds^2=-dt^2+a^2(t)\delta_{\mu\nu}dx^\mu dx^\nu\label{FRW metric}
\end{equation}
$a(t)$ being the scale factor. With this assumption the non-metricity scalar reduces to $Q=6H^2$, $H=\frac{\dot{a}}{a}$ 
being the Hubble function \cite{Dutta 2024}. The upper dot here denotes derivative with respect to t. \\
Now , if the energy momentum tensor is taken to be in the form of that of a perfect fluid, i.e 
\begin{equation}
    T_{\mu\nu}=(\rho_m+p_m)u_\mu u_\nu+p_mg_{\mu\nu}\label{Perfect fluid}
\end{equation}
where $p_m$  and $\rho_m$ represent respectively the pressure and the energy density of matter, then from (\ref{Einstein Field eq}) one can derive the corresponding modified FLRW equations for $f(Q)$ gravity as follows :

\begin{eqnarray}
    3H^2&=&\frac{1}{2f_Q}\left(-\rho_m+\frac{f}{2}\right)\label{FRW first}\\
    \dot{H}+3H^2+\frac{\dot{f}_Q}{f_Q}H&=&\frac{1}{2f_Q}\left(p_m+\frac{f}{2}\right)\label{FRW}
\end{eqnarray}
Now, to have a analytic form of  theoretically predicted H(z), after some direct calculations we rewrite the FRW equations  (\ref{FRW first}) and (\ref{FRW}) as follows:
\begin{eqnarray}
H^2 &=& \frac{1}{12f_Q}\left(-Q\Omega_m+f\right)\label{FRW Modified1}\\
\dot{H}&=&\frac{1}{4f_Q}\left(Q\Omega_m-4H\dot{Q}f_{QQ}\right)\label{FRW Modified2}\
\end{eqnarray}
Here $\Omega_{m}=\frac{\rho_{m}}{3H^2}$ is matter density parameter. Also, the matter content of the universe satisfies the following continuity equation:
\begin{equation}\label{Cont}
\dot{\rho_m}+3H(\rho_m+p_m)=0
\end{equation}
For a given form of $f(Q)$ from above three equations we can derive the analytic form of $H(z)$.

\section{Observational Analysis}
In this section $f(Q)$ gravity model is examined in view of recent observations. In particular we have used three sets of data viz. 31-point Hubble, Pantheon type SNIa data and combined Hubble+Pntheon data for our purpose.
Hubble dataset is publicly available from \url{https://github.com/Ahmadmehrabi/Cosmic_chronometer_data and} and Pantheon is from \url{https://github.com/PantheonPlusSH0ES/DataRelease/tree/main/Pantheon%2B_Data/1_DATA/header_overrides} \cite{NeurIPS 2024}.

Using  Markov Chain Monte Carlo (MCMC) method we estimate best fit values of the model parameters. The MCMC method enables reliable estimation of model parameters consistent with observational data. The primary aim of this technique is to maximize the likelihood function given below

\begin{equation}
L_{\mathrm{tot}} = e^{-\chi^{2}/2},
\end{equation}
Alternatively we can minimize the total chi-square function \cite{Dutta 2024}.

The total chi-square is written as
\begin{equation}
\chi^{2} = \chi^{2}_{H(z)} + \chi^{2}_{SN},
\end{equation}
where $\chi^{2}_{H(z)}$ and $\chi^{2}_{SN}$ denote the contributions from the $H(z)$ and Pantheon+ datasets, respectively. This method enables reliable estimation of model parameters consistent with observational data.

\subsection*{$H(z)$ Data}
A set of 31 correlated measurements of the Hubble parameter is available within the redshift range \[0.07 \leq z \leq 1.965\] \cite{Dutta 2024}. The chi-square function for the Hubble dataset is given by
\begin{equation}
\chi^{2}_{H(z)} = \sum_{i=1}^{N} \frac{\left[H_{\mathrm{obs}}(z_i)-H_{\mathrm{th}}(z_i)\right]^2}{\sigma^2_H(z_i)},
\end{equation}
where $H_{\mathrm{obs}}(z_i)$ and $H_{\mathrm{th}}(z_i)$ respectively represents the observed  and theoreticaly predicted value of Hubble parameter and $\sigma_H(z_i)$ denotes the corresponding uncertainty.
\subsection*{Pantheon+ Data}
The Pantheon+ dataset consists of 1048 Type Ia supernovae spanning the redshift range \[0.00122 \leq z \leq 2.2613\]. The model parameters are constrained by comparing observed and theoretical distance moduli. The chi-square function for the Pantheon data set is given by
\begin{equation}
\chi^{2}_{SN}(z,p_s)= \Delta \mu_i (C^{-1}_{SN})_{ij}\Delta \mu_j ,
\end{equation}
where $p_s$ represents model parameters and $C_{SN}$ is the covariance matrix \cite{Dutta 2024}. The theoretical distance modulus is defined as
\begin{equation}
\mu(z,p_s)=5\log_{10}[d_L(z,p_s)]+\mu_0,
\end{equation}
where $\mu_0$ is a nuisance parameter and $d_L$ denotes the dimensionless luminosity distance \cite{Dutta 2024}:
\begin{equation}
d_L(z)=(1+z)\int_{0}^{z}\frac{dz'}{E(z')},
\end{equation}
with
\begin{equation}
E(z)=\frac{H(z)}{H_0}.
\end{equation}
Now, in our study we focused on a specific form of $f(Q)$ gravity : $f(Q)=Q+mQ^n$ which is called the power law model of $f(Q)$ gravity. Also, for simplicity, we assume the matter content of universe to be pressure less dust, i.e $p_m=0$.
Then using continuity equation (\ref{Cont}) for matter  along with the FRW equation (\ref{FRW Modified1}) we obtain an implicit form of Hubble function $H(z)$:\\
\begin{equation}
m\left(6H^2\right)^n(1-2n)-6H^2=6H_0^2\Omega_{0m}(1+z)^3\label{Hubble form}
\end{equation}
where $H_0$ and $\Omega_{0m}$ are respectively the current value of Hubble and matter density parameter. Also, $m$ and $n$ are not independent, rather related by the equation $m=\frac{1+\Omega_{0m}}{(1-2n)Q_0^{n-1}}$, where $Q_0$ is the present value of $Q$.\\

The best-fit values of $H_0$, $m$, and $n$ determined through the MCMC analysis, are summarized in Table-\ref{tab:bestfit}.

\begin{table}[h]
\centering
\caption{Marginalized $1\sigma$ constraints on cosmological parameters}
\begin{tabular}{lcccccc}
\hline
Dataset & $H_0$ & $\Omega_{m0}$ & $n$ & $\chi^2_{\min}$ & $\chi^2_{\min}/{\rm dof}$ \\
\hline

Hubble &
$69.00^{+1.56}_{-1.56}$ &
$0.25^{+0.044}_{-0.044}$ &
$1.257^{+0.025}_{-0.025}$ &
23.324 &
0.833 \\

Pantheon+SHOES &
$72.59^{+0.28}_{-0.28}$ &
$0.236^{+0.053}_{-0.053}$ &
$1.25^{+0.037}_{-0.037}$ &
1582.536 &
0.935 \\

Hubble+Pantheon+SHOES &
$72.67^{+0.25}_{-0.25}$ &
$0.229^{+0.035}_{-0.035}$ &
$1.260^{+0.029}_{-0.029}$ &
1720.899 &
0.995 \\

$\Lambda$CDM(Hubble+Pantheon+SHOES) &
$68.11^{+0.25}_{-0.25}$ &
$0.323^{+0.016}_{-0.016}$ &
- &
- &
- \\ 
\\

\hline
\end{tabular}
\label{tab:bestfit}
\end{table}

Since all three values of $\chi^2_{\min}$/$dof$ are very close to unity, the model provides an excellent fit to the observational data. However, recent studies have shown Hubble tension, between the measurements of the Hubble parameter from the Planck satellite data and other methods used to study the universe. The Planck collaboration estimated the Hubble parameter as 
$H_0 = 67.4\pm0.5~\mathrm{km\,s^{-1}\,Mpc^{-1}}$,
while the HST team found it to be 
$H_0 = 74.03 \pm 1.42~\mathrm{km\,s^{-1}\,Mpc^{-1}}$. 
This creates a large difference, or tension between the two results. This evolution leads to our best-fit Hubble constant from the $f(Q)$ gravity model, 
$H_0 = 72.67~\mathrm{km\,s^{-1}\,Mpc^{-1}}$, 
which lies significantly closer to the local SH0ES measurement
($H_0 = 74.03 \pm 1.42~\mathrm{km\,s^{-1}\,Mpc^{-1}}$) than to the Planck CMB value ($H_0 =67.4\pm0.5~\mathrm{km\,s^{-1}\,Mpc^{-1}}$).
Consequently, the $f(Q)$ model considered in this study partially alleviates the well-known Hubble tension.

The best-fit value of $H_0$ obtained using the combined CC and
Pantheon Plus \& SH0ES datasets through the Bayesian MCMC approach lies between the Planck 2018 result ($H_0 = 67.4 \pm 0.5~\mathrm{km\,s^{-1}\,Mpc^{-1}}$) and the SH0ES
determination ($H_0 = 73.0 \pm 1.0~\mathrm{km\,s^{-1}\,Mpc^{-1}}$), suggesting a modest reduction in the Hubble tension. 

To assess the relative performance of the $f(Q)$ and $\Lambda$CDM models, we employ the Akaike Information Criterion (AIC) and the Bayesian Information Criterion (BIC). Both criteria are commonly used in statistical model selection and are briefly outlined below.
The Akaike Information Criterion is given by
\begin{equation}
\mathrm{AIC} = -2\ln L + 2k
= \chi^2_{\min} + 2k,
\end{equation}
while the Bayesian Information Criterion is

\begin{equation}
\mathrm{BIC} = -2\ln L + k\ln N
= \chi^2_{\min} + k\ln N.
\end{equation}

In this expression,

\begin{equation}
L = \exp\left(-\frac{\chi^2_{\min}}{2}\right)
\end{equation}

denotes the maximum likelihood, with $k$ being the number of free parameters in the model and the total number of data points employed in the analysis is represented by $N$. Taking the standard $\lambda$CDM cosmology as the reference model, the relative performance of any given model can be assessed by calculating the difference 
\begin{equation}
\Delta X = X_{\mathrm{Model}} - X_{\Lambda\mathrm{CDM}},
\end{equation}
where $X$ stands for either the AIC or BIC x\cite{NeurIPS 2024}. The resulting value can then be used to interpret the model's goodness of fit. In the case of AIC, a good fit is indicated when $\triangle AIC\le 2$, a moderate fit when $4\le\triangle AIC\le 7$ and a poor fit when  $\triangle AIC\ge 10$. For BIC, the evidence against the model is regarded as positive when  $2\le\triangle BIC\le 6$, strong when  $6\le\triangle BIC\le 10$ and very strong when $\triangle BIC> 10 2$. The computed values of AIC, BIC and $\chi^2$ for our model in case of three different data sets are given in Table~\ref{tab:results}. It compares them to the traditional $\Lambda$CDM cosmology using observational datasets: CC, PPS, and CC+PPS. The relative differences between $\Delta$AIC and $\Delta$BIC in relation to the $\Lambda$CDM model are also accounted for to enable a fair comparison. 

\begin{table}[htbp]
\centering
\caption{Calculated values of $\chi^2$, AIC, and BIC for the $f(Q)$ and
$\Lambda$CDM models.}
\label{tab:results}
\begin{tabular}{llccc}
\toprule
Dataset & Model & $\chi^2$ & AIC & BIC \\
\midrule
Hubble, 31 points & $f(Q)$ & 23.324 & 29.324 & 33.626 \\
                  & $\Lambda$CDM & 21.844 & 25.844 & 28.712 \\
\midrule
Pantheon+SH0ES, 1701 points & $f(Q)$ & 1582.536 & 1588.536 & 1604.853 \\
                            & $\Lambda$CDM & 1580.756 & 1584.756 & 1595.634 \\
\midrule
Hubble + Pantheon+SH0ES & $f(Q)$ & 1720.899 & 1726.899 & 1743.270 \\
                        & $\Lambda$CDM & 1720.439 & 1724.439 & 1735.353 \\
\bottomrule
\end{tabular}
\end{table}

\begin{table}[htbp]
\centering
\caption{Relative differences in AIC and BIC with $\Lambda$CDM baseline.}
\label{tab:delta}
\begin{tabular}{lccc}
\toprule
Dataset & $\Delta\chi^2$ & $\Delta$AIC & $\Delta$BIC \\
\midrule
Hubble, 31 points & 1.48 & 3.48 & 4.914 \\
Pantheon+SH0ES, 1701 points & 1.78 & 3.78 & 9.219 \\
Hubble + Pantheon+SH0ES & 0.46 & 2.46 & 7.917 \\
\bottomrule
\end{tabular}
\end{table}

\begin{figure}
        \centering
        \includegraphics[width=0.75\linewidth]{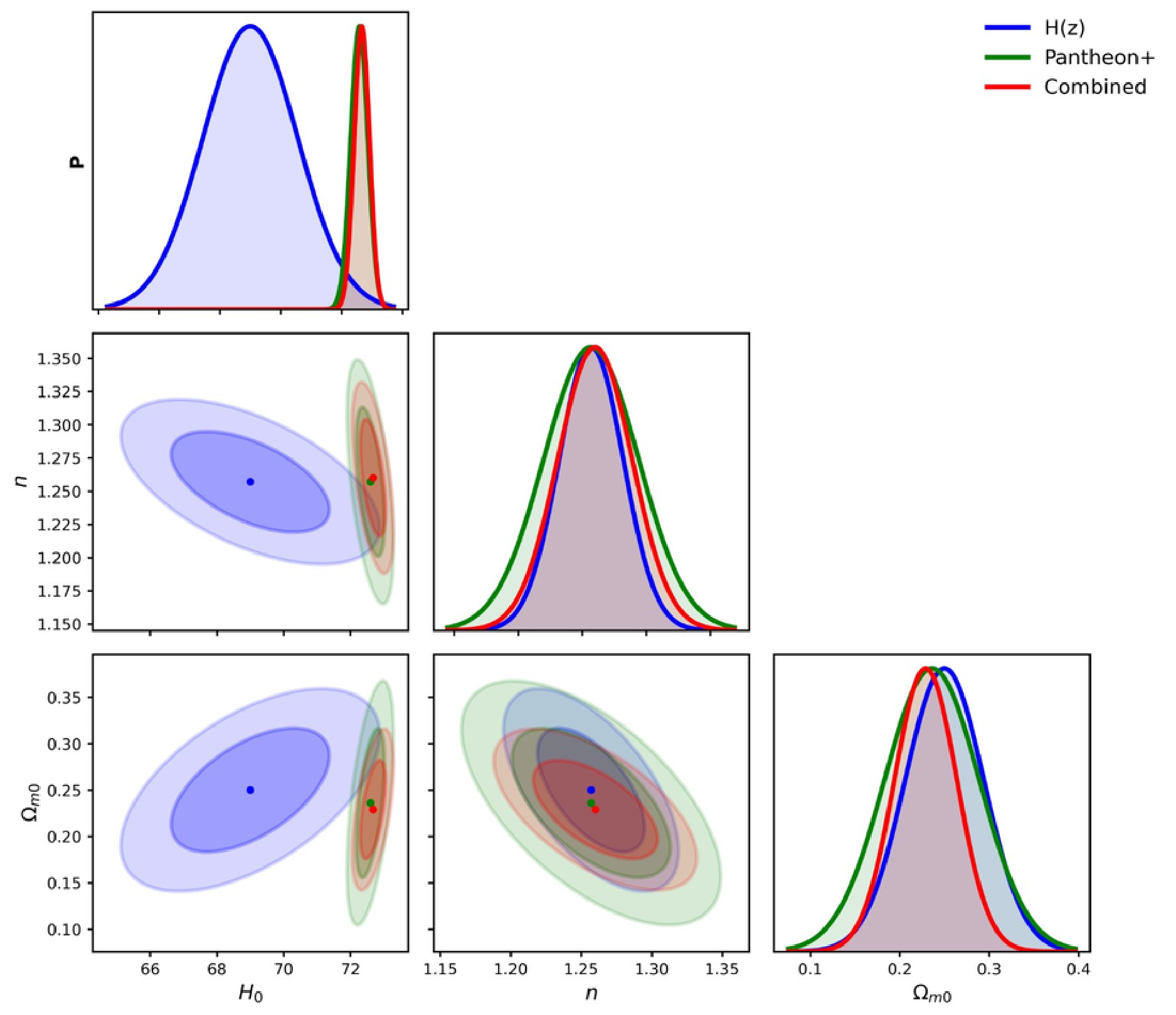}
        \caption{ MCMC $1\sigma$ and $2\sigma$ confidence contour plot obtained from $H(z)$ and Pantheon+ data set}
    \label{contour}
\end{figure}

The values of $\Delta$AIC and $\Delta$BIC clearly indicate that the results are statistically consistent enough and in full agreement with the observational data.\\ However, to understand correlations between the model parameters and their influence over evolution dynamics of universe we draw contour plots shown in Figure(\ref{contour}). It shows the confidence contours along with marginalized posterior distributions derived from the combined $H(z)$ and Pantheon+ datasets.

\begin{figure}
                    \centering
                    \includegraphics[width=0.75\linewidth]{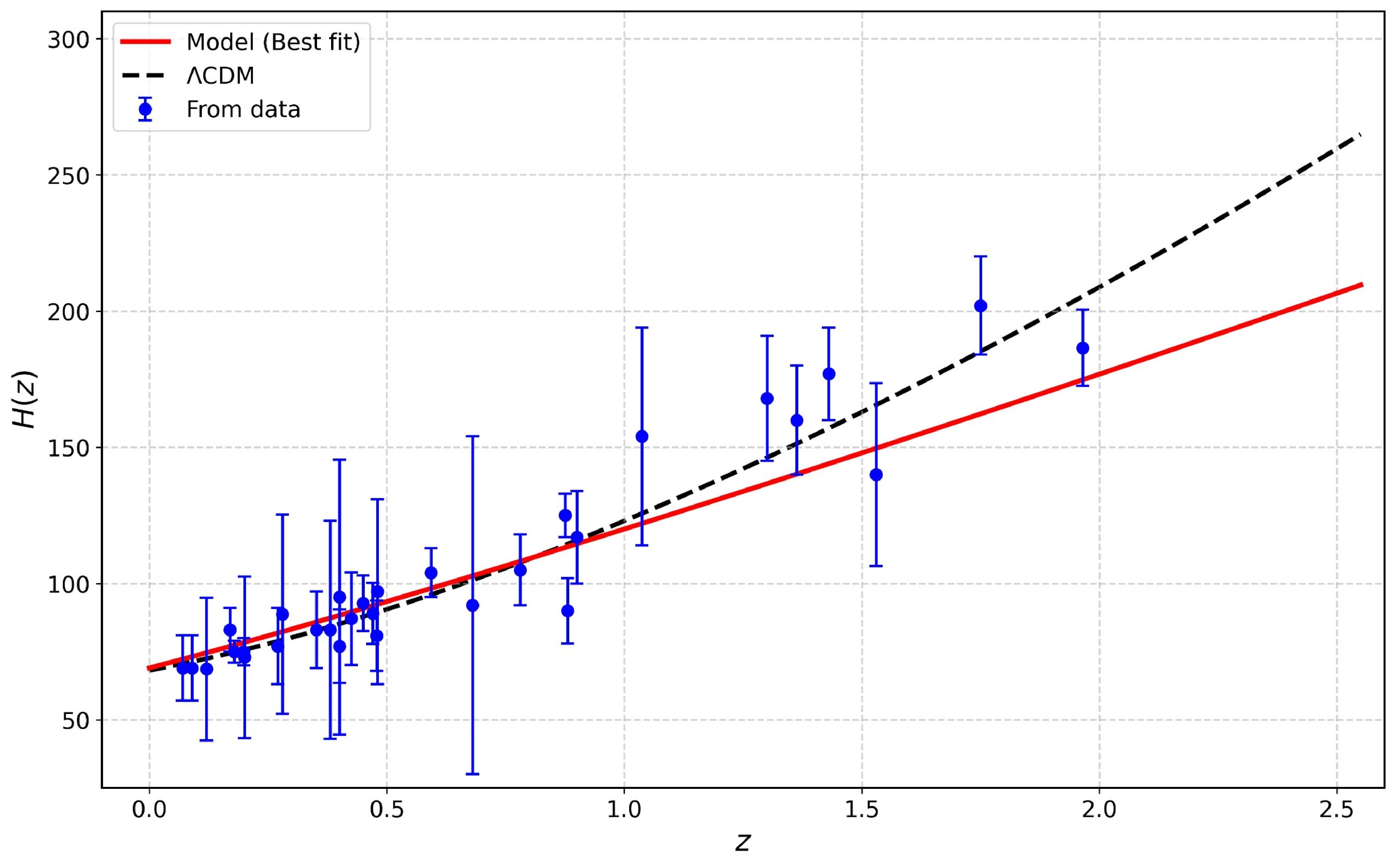}
                    \caption{$H(z)$ in redshift from 31 data points of $H(z)$ data (upper panel). Blue error bars from the data set, red solid line for the model and broken black line for $\Lambda$CDM}
    \label{error bar}
\end{figure}

\section*{Contour plot Analysis}

In Figure-\ref{contour},the blue line represents Cosmic Chronometers or Hubble parameter data. This typically has the widest distribution, indicating higher uncertainty. The green line represents Pantheon Type Ia Supernovae data. The red line represents the joint analysis of both datasets. As expected, the red contours are the smallest and the peaks are the sharpest, showing how combining data reduces overall uncertainty. The parameter $n$ controls the power-law behavior of the modified gravity term. One of the most striking features of the corner plot is the extremely tight constraint on $n$. All datasets strongly favor values close to unity, with the combined best-fit centered around 1.26.
This result is cosmologically important because $n=1$ corresponds to behavior very close to the standard cosmological scenario. The narrowness of the combined distribution suggests that the model is highly sensitive to changes in $n$, and even small departures from unity significantly affect the expansion history. The matter density parameter $\Omega$ is also tightly constrained. The combined dataset yields a sharply peaked distribution around 0.229.

The contour plots involving $\Omega$ show relatively weak correlation with the Hubble parameter and negative correlation with $n$. Weak parameter correlations imply that the matter density can be estimated relatively independently of the modified gravity parameters, increasing the robustness of the cosmological interpretation. In the contours involving $H$ and $\Omega$, no tilt may be observed in combined dataset but a slight positive tilt is observed in Hubble dataset. No tilt in the combined dataset indicates that the Hubble constant and the exponent $n$ are almost statistically independent.However mild negative tilt is observed in Hubble data. The observational data constrain these parameters separately with little degeneracy.

Figure -\ref{error bar} presents an error bar plot which shows the evolution of the Hubble parameter $H(z)$ over the change of redshift $z$ comparing observational data against the theoretical predictions of both the modified $f(Q)$ gravity model and the standard $\Lambda$CDM cosmology.
The blue points with vertical error bars indicate the observational Hubble data. The error bars indicate the observational uncertainties associated with each measurement. At low redshift, the uncertainties are relatively small, while at higher redshift the errors become significantly larger, reflecting the difficulty of measuring the expansion rate in the distant universe.

At low redshift, both theoretical curves closely follow the observational data, suggesting that the modified gravity model accurately captures the universe's late-time expansion history. The difference between the two models in this region is very small, showing that the $f(Q)$ model behaves similarly to $\Lambda$CDM near the present epoch.As the redshift increases, the deviation between the two theoretical curves becomes more noticeable. The $\Lambda$CDM curve predicts a slightly faster growth of $H(z)$ compared with the modified $f(Q)$ model. The red best-fit curve remains systematically below the dashed black curve at intermediate and high redshifts. This behavior suggests that the modified gravity model predicts a comparatively slower expansion rate in the earlier universe.
Most of the observational points remain consistent with both theoretical curves within their error limits. However, the large uncertainties at high redshift prevent a strong discrimination between the models using Hubble data alone. This explains why combining Hubble data with Pantheon+ supernova observations is important for obtaining tighter constraints on the cosmological parameters.

\section{Thermodynamic Analysis}
In this section we have studied our universe in $f(Q)$ gravity framework from thermodynamic point of  view. We consider our universe to be an isolated thermodynamical system bounded by apparent horizon. Since there is a continuous energy flow between the system and bounding horizon, instead of equilibrium thermodynamics a non-equilibrium thermodynamic technique is more suitable to be applied. In particular we have checked validity of Generalized Second Law of Thermodynamics (GSLT) and Thermodynamic Equilibrium(TE) for our model and found some constraints on the model parameters from thermodynamic point of view \cite{Biswas 2019}.
\subsection{Unified Law of Thermodynamics in f(Q) gravity}
First we shall briefly discuss the Unified law of thermodynamics \cite{Mitra 2015} for a two dimensional spherical space time in Einstein gravity and then will apply it to find horizon entropy in $f(Q)$ gravity.\\
The metric given in Eq. (\ref{FRW metric}), describing a spatially flat, homogeneous, and isotropic FLRW universe, can be recast as
\begin{equation}
    ds^2=h_{ab}dx^adx^b+R^2d\Omega_2^2
\end{equation}
where $R = ar$ denotes the area radius and $h_{ab}=diagnal\left(-1,a^2\right)$ represents the metric on 2-space given by $\left(x^0=t,x^1=r\right)$. Now, upon introducing double null co-ordinates $\xi^{\pm}$ the above line element reduces to the form \cite{Mitra 2015}
\begin{equation}
     ds^2=-2d\xi^+d\xi^-+R^2d\Omega_2^2\label{metric double null form}
\end{equation}
with
\begin{equation}
 \delta_\pm=\frac{\partial}{\partial\xi^\pm}=-\sqrt{2}\left(\frac{\partial}{\partial t}\mp\frac{1}{a}\frac{\partial}{\partial r}\right)    
\end{equation}
as the future pointing null vectors.\\

Generally, the energy supply vector $\psi$ and the work density $W$ are  respectively given by 
\begin{eqnarray}
    \psi_a &=& T^a_b\partial_bR + W\partial_aR\label{energy supply}\\
 and~~   W &=&-\frac{1}{2}T^{ab}T_{ab}\label{work density}
\end{eqnarray}
The Einstein's field equation can be written in the form of\textbf{ Unified First Law}\cite{Hayward 1998} considering only the $(0,0)$ component from the energy supply and work density vector equations given by equations(\ref{energy supply}) and (\ref{work density}) as follows:
\begin{equation}
 dE=A\psi+WdV
\end{equation}
Where $A=4\pi R_A^2$ and $V=\frac{4}{3}\pi R_A^3$ are respectively the area of the bounding apparent horizon and volume of the sphere bounded by the horizon, $R_A$ being the radius of apparent horizon.
Now, if we write $f(Q)$ as $f(Q)=Q+F(Q)$ the  FLRW equations can be written as
\begin{equation}
3H^2=\rho+\rho_e\label{FRW1}
\end{equation} 

\begin{equation}
\dot{H}=-\frac{1}{2}(\rho+p)-\frac{1}{2}(\rho_e+p_e)\label{FRW2}    
\end{equation}
where 
\begin{eqnarray}
 \rho_e&=&Q(1+F_Q)-\frac{F}{2}\\
 \rho_e+p_e &=& -2\dot{H}\left(2+F_Q+2QF_{QQ}\right)\label{curvature pressure and density}
\end{eqnarray}

For our present model, using the above definitions, work density and energy supply vector can be represented as
\begin{equation}
W=-\frac{1}{2}(\rho-p)+\frac{1}{2}(\rho_e-p_e)
\end{equation}
and
\begin{equation}
\psi=\psi_m+\psi_e
\end{equation}
with 
\begin{equation}
\psi_m=-\frac{1}{2}(\rho+p)HRdt+\frac{1}{2}(\rho+p)adr
\end{equation}
and 
\begin{equation}
  \psi_e=  -\frac{1}{2}(\rho_e+p_e)HRdt+\frac{1}{2}(\rho_e+p_e)adr
\end{equation}
Now, only the pure matter energy supply $A\psi_m$ contributes to the heat flow $\delta Q$ in the Clausius relation $\delta Q=T\delta S$.

Using the double null vector $\partial_\pm$ as the basis, any vector $\xi$ tangential to the apparent horizon can be expressed as
\begin{equation}
\xi=\xi_+\partial_++\xi_-\partial_-
\end{equation}
where the ratio of the coefficients can be derived from the fact that $\partial_+R_A=0$, $R_A$ being a trapping horizon. So, in $(t,r)$ coordinates $\xi$ takes the following form :
\begin{equation}
    \xi =\frac{\partial}{\partial t}-(1-2\epsilon)Hr\frac{\partial}{\partial r}
\end{equation}
where $\epsilon=\frac{\dot{R_A}}{2}$. Thus projecting the Unified first law along $\xi$, the first law of thermodynamics appears as:
 \begin{equation}
<dE,\xi>=\frac{\kappa}{8\pi G}<dA,\xi>+<WdV,\xi>or
\end{equation}
Consequently one can write the Clausius relation as
\begin{equation}
\delta Q=T<dS,\xi>=<A\psi_m,\xi>=<dE,\xi>=\frac{\kappa}{8\pi G}<dA,\xi>-<A\psi_e,\xi>
\end{equation}
or in explicit form, 
\begin{equation}
\delta Q=<A\psi_m,\xi>=-\frac{2\epsilon(1-\epsilon)}{G}+A(1-\epsilon)(\rho_e+p_e)\label{explicit UFL}
\end{equation}
The Hawking temperature on apparent horizon is given by 
\begin{equation}
T_A=\frac{\kappa_A}{2\pi}=\frac{(1-\epsilon)}{2\pi R_A}
\end{equation}
Then using (\ref{explicit UFL}) we obtain
\begin{equation}
    dS_A=d\left(\frac{A}{4G}\right)-\frac{\pi R_A^3}{G}(\rho_e+p_e)dt
\end{equation}
which on integration the final form of the entropy on apparent horizon is obtained as follows :
\begin{equation}
    S_A=\left(\frac{A}{4G}\right)-\int\frac{\pi R_A^3}{G}(\rho_e+p_e)dt\label{horizon entropy equation}
\end{equation}
Now, using the above expression for horizon entropy and equation (\ref{curvature pressure and density}), after some simple calculations one can derive the following $S_A$ in $f(Q)$ gravity :

\begin{equation}
S_A=\frac{A}{4G}+\frac{6\pi}{G}\int (\frac{2+F_Q+2QF_{QQ}}{Q^2})\dot{Q}dt \label{horizon entropy f(Q)}
\end{equation}
\subsection{Theory of irreversible
thermodynamics in f(Q) gravity}
In this work we have considered the first order Eckart theory of irreversible thermodynamics \cite{Gang 2009, Biswas 2013}. In the theory of of irreversible thermodynamics the Clausius relation 
\begin{equation}
 dS=\frac{\delta Q}{T}\label{Clausius}   
\end{equation}
is modified as the following entropy balance equation:
\begin{equation}
    dS_T=\frac{\delta Q}{T}+d_iS=d_eS+d_iS\label{entropy balance}
\end{equation}

where the term $d_is$ occurs in the equation due to internal production process. For irreversible process, $d_is>0$ and zero otherwise. In classical sense $\delta Q$ denotes the heat exchange between the system and its surrounding where $d_iS$ represents the entropy produced internally from the uncompensated heat associated with an irreversible process in the system.\\
Let the internal entropy production density be represented by $\sigma$ and entropy flow density be represented by $\vec{J_s}$. Then if  local equilibrium is assumed, one can write 
\begin{equation}
    \frac{d_eS}{dt}=-\int_\Sigma \vec{J_s}.d\Sigma
\end{equation}
\begin{equation}
    \frac{d_iS}{dt}=\int_V \sigma dV\label{internal production}
\end{equation}
where $\Sigma$ is the surface of the horizon and $V$ is the volume inside the bounding horizon.
Now, internal entropy production and entropy flow in general can arise from a variety of processes such as convection, heat conduction, diffusion, and others. For simplicity, however, we restrict our consideration to heat conduction as the dominant contributing mechanism. Then one can write

\begin{equation}
    \frac{d_eS}{dt}=\frac{A\vec{J_q}}{T}\label{heat current entropy}
\end{equation}
\begin{equation}
    \frac{d_iS}{dt}=\sigma.V\label{irreversible entropy equation}
\end{equation}
where $\vec{J_q}$ and $T$ denote respectively the heat current and the temperature of the system and $A$ represents the bounding horizon's surface area. $\vec{J_q}$ and $\sigma$ are related by the following equations:
\begin{equation}
    \vec{J_s}=\frac{\vec{J_q}}{T}
\end{equation}
\begin{equation}
    \sigma=\vec{J_q}.\nabla\left(\frac{1}{T}\right)\label{entropy production}
\end{equation}
Now if the universe is assumed to be a Bekenstein system bounded by apparent horizon, then the horizon entropy in $f(Q)$ gravity can be evaluated using Unified law of thermodynamics from eqn. no.(\ref{horizon entropy f(Q)}). Thus we have 
\begin{equation}
  \frac{d_eS}{dt}=\frac{\dot{A}}{4G}+\frac{6\pi \dot{Q}}{G}\left(\frac{2+F_Q+2QF_{QQ}}{Q^2}\right)\label{horizon entropy equation 2}   
\end{equation}
Comparing the above equation with eqn.(\ref{heat current entropy}), it can be derived that
\begin{equation}
    |\vec{J_q}|= \frac{T\dot{R_A}}{2R_AG}\left(3+F_Q+2QF_{QQ}\right)
\end{equation}
Now, following Eckart–Fourier law which provides relationship between heat flux and the temperature of a fluid system, we can write 
\begin{eqnarray}
  \vec{J_q}=-\lambda \vec{\nabla}T  
\end{eqnarray}
The above law states that a temperature gradient gives rise to an energy flux and here $\lambda>0$ denotes the thermal conductivity. Now, substituting this value of $|\vec{J_q}|$ from the above Fourier law in the equation (\ref{entropy production}) for $\sigma$ one can derive:  
\begin{eqnarray}
 \sigma&=&\frac{|J_q|^2}{\lambda T^2}\\
 ~&=&\frac{\dot{R_A}^2}{4\lambda R_A^2G^2}\left(3+F_Q+2QF_{QQ}\right)^2\label{entropy production2}
\end{eqnarray}
Using Equations (\ref{irreversible entropy equation}), (\ref{horizon entropy equation 2}) and (\ref{entropy production2}), finally the total entropy change with respect to time is obtained as
\begin{equation}
   \frac{dS_T}{dt}=\frac{2\pi R_A \dot{R_A}X}{G}(1+\frac{X}{6\lambda G})\label{entropy}
\end{equation}
Where, $X=3+F_Q+2QF_{QQ}$.\\
The expression for total entropy clearly shows it's dependence  on the non equilibrium factor $\lambda$.

\subsection{Generalized second law of thermodynamics and thermodynamic equilibrium}
An isolated macroscopic system always evolves towards thermodynamic equilibrium i.e to a state of maximum entropy and therefore entropy of such a system should always increase. Thus, the generalized second
law of thermodynamics (GSLT) and thermodynamical equilibrium (TE) hold for a matter filled universe bounded by a horizon, if the following conditions are hold:
\begin{itemize}
    \item $\frac{dS_T}{dt}\geq 0$ (Condition for GSLT
    \item $\frac{d^2S}{dt^2}<0$ (Condition for TE)
\end{itemize}
From eqn.(\ref{entropy}) it can be inferred that compliance with either of the following conditions results in validity of GSLT:
\begin{table}[htb]
    \centering
    \caption{Conditions for satisfying GSLT}\label{Table1}
    \begin{tabular}{|c|c|} 
    \hline
    \textbf{Cases} &  \textbf{Condition}\\
    \hline
       $\dot{R_A}>0$ (quintessence era) & Either $X \geq 0$ or $X < -6\lambda G$  \\ \hline
      $\dot{R_A}<0$ (Phantom era) & $-6\lambda G<X<0$\\ \hline
    \end{tabular}
\end{table}

Now,
\begin{equation}
   \frac{d^2S_T}{dt^2}=\frac{dS_T}{dt}(Y+\frac{2\dot{X}}{X})-\frac{2\pi R_A\dot{R_A}\dot{X}}{ G}\label{TE}
\end{equation}
where $Y=\frac{\ddot{R_A}}{\dot{R_A}}+\frac{\dot{R_A}}{R_A}=\left(3q-\frac{j-1}{1+q}\right)H$ and $q=-1-\frac{\dot{H}}{H^2}$ and $j=\frac{\ddot{H}}{H^3}-3q-2$ respectively being the deceleration and jerk parameter. Assuming validity of GSLT i.e considering $\frac{dS_T}{dt}\geq 0$, we derive conditions for the validity of TE in both quintessence and phantom era. These findings are summarized in table-\ref{Table3}.

% Requires: \usepackage{multirow}
\begin{table}[h]
    \centering
          \caption{Conditions for Thermal equilibrium based on $\frac{dS_T}{dt}\geq 0$ }\label{Table3}
    \begin{tabular}{|c|c|c|}
    \hline
    \textbf{Cases} & \textbf{Condition} &  \textbf{Conclusion}\\
    \hline
    \multirow{4}{4cm}{$\dot{R_A}>0$} & $\dot{X}>0, Y+\frac{2\dot{X}}{X}>0$ & TE holds if $\frac{dS_T}{dt}<\frac{2\pi R_A\dot{R_A}\dot{X}}{G\left(Y+\frac{2\dot{X}}{X}\right)}$ \\
   
   & $\dot{X}>0, Y+\frac{2\dot{X}}{X}<0$ & TE holds\\
    
    (Quintessence era) & $\dot{X}<0, Y+\frac{2\dot{X}}{X}>0$ & TE does not hold\\
    
    & $\dot{X}<0, Y+\frac{2\dot{X}}{X}<0$ & TE holds if $\frac{dS_T}{dt}>\frac{2\pi R_A\dot{R_A}|\dot{X}|}{G|Y+\frac{2\dot{X}}{X}|}$\\
    \hline\hline
     \multirow{4}{4cm}{$\dot{R_A}<0$} & $\dot{X}>0, Y+\frac{2\dot{X}}{X}>0$ & TE does not hold \\
   
   & $\dot{X}>0, Y+\frac{2\dot{X}}{X}<0$ & TE holds if  $\frac{dS_T}{dt}>\frac{2\pi R_A
   |\dot{R_A}||\dot{X}|}{G|Y+\frac{2\dot{X}}{X}|}$ \\
    
    (Phantom era) & $\dot{X}<0, Y+\frac{2\dot{X}}{X}>0$ & TE holds if $\frac{dS_T}{dt}<\frac{2\pi R_A|\dot{R_A}||\dot{X}|}{G\left(Y+\frac{2\dot{X}}{X}\right)}$ \\
    
    & $\dot{X}<0, Y+\frac{2\dot{X}}{X}<0$ & TE holds \\
    \hline
   \end{tabular} 
\end{table}
Now, applying results obtained from above table,  we try to find some feasible limits of the model parameters for power law model of $f(Q)$ gravity. Since in this model $f(Q)=Q+mQ^n$, we identify $F(Q)=mQ^n$. Consequently,$ F_Q, F_{QQ}$ are calculated as follows:\\

$F_Q=mnQ^{n-1}$\\
$F_{QQ}=mn(n-1)Q^{n-2}$\\
Then we derive $X=3+mn(2n-1)Q^{n-1}$ and $\dot{X}=-2(n-1)(X-3)H(1+q)$\\\\

In Table-\ref{Table4} we show the restrictions on the parameters $m$ and $n$ for validity of GSLT and TE in both quintessence and phantom era.
\begin{table}[h]
    \centering
     \caption{Conditions for Thermal equilibrium based on $\frac{dS_T}{dt}\geq 0 $ }\label{Table4}
    \begin{tabular}{|c|c|c|}
    \hline
    \textbf{Cases} & \textbf{Condition} &  \textbf{Conclusion}\\
    \hline
    \multirow{4}{4em}{$\dot{R_A}>0$} &$\frac{1}{2} < n < 1, m < 0$ &  TE  hold\\
   
   & $ \frac{1}{2}<n<1,m>0$ & TE hold\\
    
    (Quintessence era) & $0<n<\frac{1}{2},m>0$ & TE hold\\
    
    & $0<n<\frac{1}{2},m<0$ & TE holds\\
    & $n>1,m<0$ & TE may or may not hold\\
    \hline\hline
     \multirow{4}{4em}{$\dot{R_A}<0$} & $\frac{1}{2} < n < 1, m < 0$ &  TE  hold  \\
   
   & $0<n<\frac{1}{2},m>0$ & TE holds \\
    
    (Phantom era) & $n>1,m<0$ & TE does not hold \\
    \hline
   \end{tabular} 
\end{table}
We see from this table that when $n>1$ and $m<0$, GSLT holds in both quintessence and phantom era but TE does not hold in phantom era and may or may not hold in quintessence era. Also, in quintessence and phantom era $m,n$ should satisfy the inequalities  $m>-\frac{3}{n(2n-1)Q^{n-1}}$ respectively and  $m<-\frac{3}{n(2n-1)Q^{n-1}}$ i,e choice of $m$ depends on $n$ and  $Q$. Here, it is to be noted that there is a continuous energy (in current work, heat) exchange  between the system and it's surroundings which hinders to attain thermal equilibrium i.e maximum entropy ever. It follows, then, that thermodynamic equilibrium cannot be maintained when a system undergoes an irreversible process. From this point of view we are interested to find that range of $m$ and $n$ for which GSLT will hold but TE will not hold. Now, in case of quintessence era we observe from Table-\ref{Table4} that except for the combination $m<0,n>1$,TE hold for any other combination of $m.n$. Since within this range of $m,n$ TE may or may not hold, our goal is to find more specific range for $m,n$ where TE will not hold. However, since $m=\frac{1+\Omega_{0m}}{(1-2n)Q_0^{n-1}}$, for $n>1$ the obvious consequence is $m<0$. In quintessence era when $m<0,n>1$ we have $0<X<3$ and it is possible only when $n$ lies in a very small neighourhood of $n$. Since $m$ is expressed in terms of $n$, we focus on finding the specific range of $n$ only. For this purpose, we check for which range of values of $n$, both $\frac{d^2S_T}{dt^2}$ and $\frac{dS_T}{dt}$ are positive. 
In case of phantom region from  Table\ref{Table3} we find that for $m\leq-1$ and $n>1.1$ with $m<-\frac{3}{n(2n-1)Q^{n-1}}$ GSLT is valid but TE does not satisfy. \\
To get more insights about $n$ we plot 3D graphs for $\frac{dS_T}{dt}$ and $\frac{d^2S_T}{dt^2}$ against $n$ and $\lambda$ for present time ($z=0$) considering current values of Hubble and matter density parameter obtained from our observational analysis in previous section. To be specific we considered here the values obtained using combined $H(z)$+Pantheon+ data sets.

\begin{figure}[htbp]
    \centering

    \includegraphics[width=0.85\textwidth]{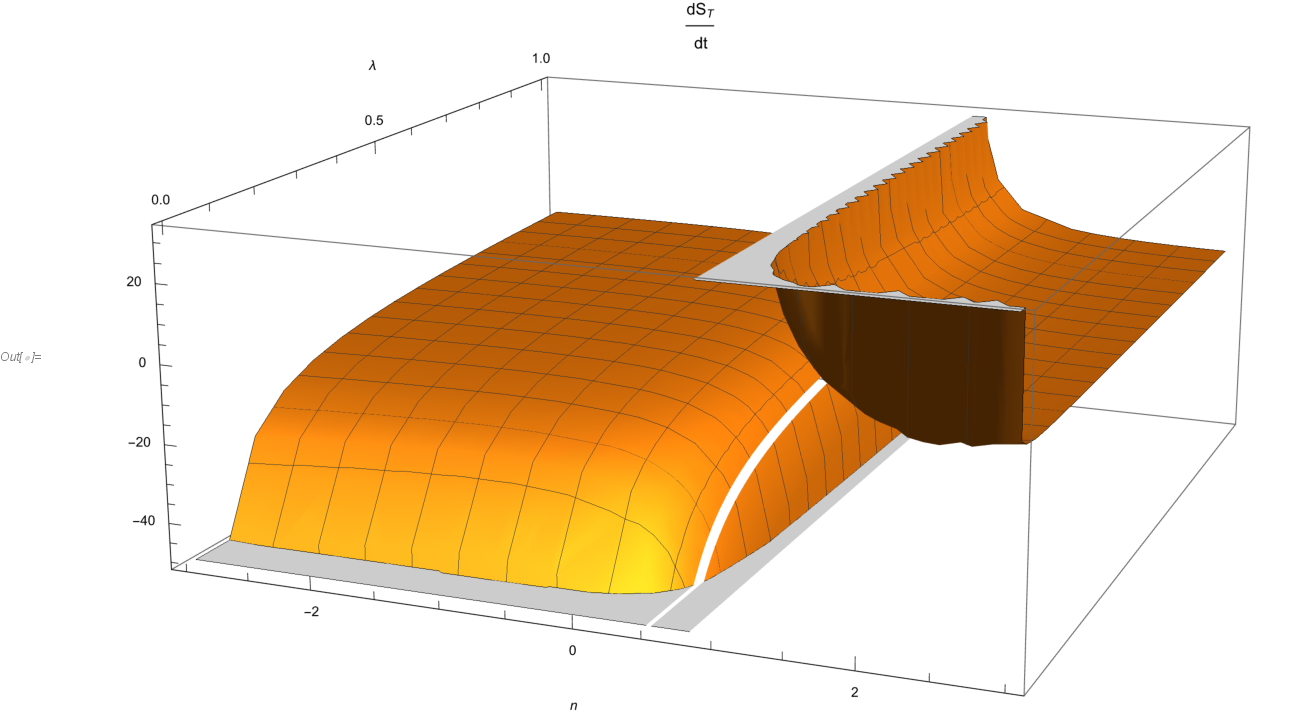}
    \caption*{\textbf{Figure 4(a)}: 3D plot for $\frac{dS_T}{dt}$ for $0<n<3$ and $0<\lambda<1$}
    \label{fig:figure1}

    \vspace{0.5cm}

    \includegraphics[width=0.70\textwidth]{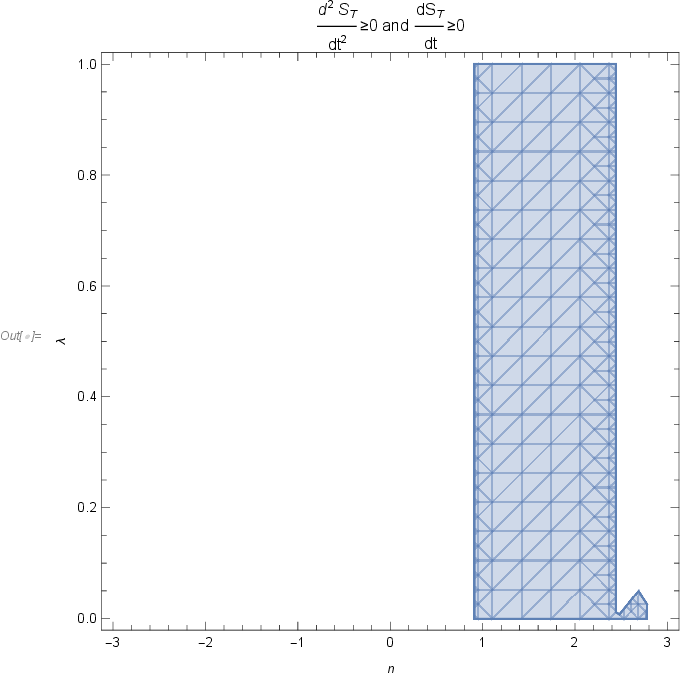}
     \caption*{\textbf{Figure 4(b)}}
    \label{fig:figure2}

    \vspace{0.5cm}

    \includegraphics[width=0.85\textwidth]{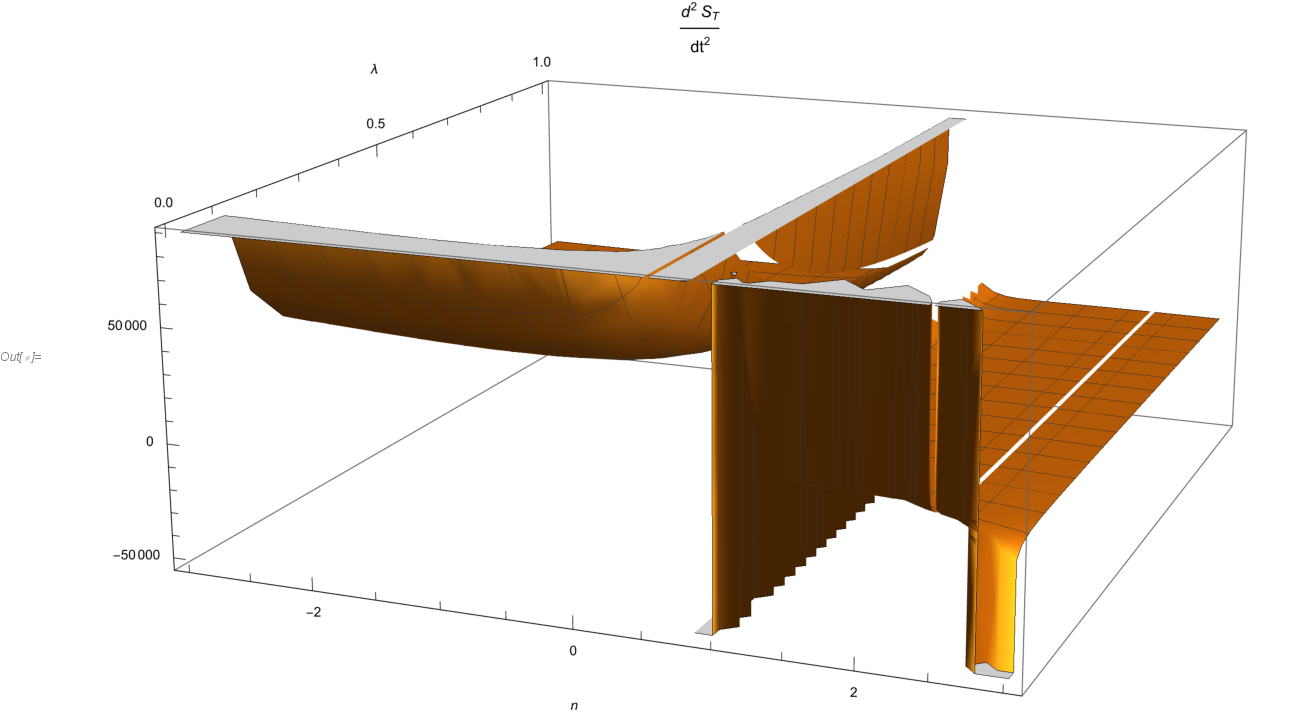}
    \caption*{\textbf{Figure 4(c)}: 3D plot for $\frac{d^2S_T}{dt^2}$ for $0<n<3$ and $0<\lambda<1$}
    \label{fig:figure3}

\end{figure}
In fig 4(a) and 4(b) we observe that for $0\leq\lambda\leq 1$ and $n\le 1$, though  $\frac{d^2S_T}{dt^2}\geq 0$ , but $\frac{dS_T}{dt}$ is not positive. $\frac{dS_T}{dt}$  is positive only when $n\geq 1$. It is clearly understandable from fig 4(c) where from the region plot we see that for $1\leq n\leq 2.4$ both $\frac{dS_T}{dt}$ and $\frac{d^2S_T}{dt^2}$ positive for entire range of $\lambda\in(0,1$.

 \section{Behaviour of Cosmographic parameters within acceptable range of Model specific parameters }
 Now we check the behavior of various cosmographic parameters viz $q$(deceleration), $j$(jerk) and $s$(snap) within  specific range of $m,n$ that are compatible with the results obtained in statistical and thermodynamic analysis. The cosmographic parameters  are defined as follows:
\begin{eqnarray}
    q(t)&=&-1-\frac{\dot{H}}{H^2}\\
    j(t)&=&-(3q+2)+\frac{\ddot{H}}{H^3}\\
    s(t)&=& 4j + 3q(q + 4) + 6+\frac{\dddot{H}}{H^4}
\end{eqnarray}
In $f(Q)$ gravity the above cosmographic parameters with the use of FRW equations (\ref{FRW Modified1}) and (\ref{FRW Modified2}) take the following forms :
\begin{eqnarray}
    q&=&-1-\frac{3\Omega_m}{2\left(f_Q+2Qf_{QQ}\right)}\\
    j&=&1-\frac{9}{2}\frac{Q\Omega_m^2\left(3f_{QQ}+2Qf_{QQQ}\right)}{\left(f_Q+2Qf_{QQ}\right)^3}\label{decl}\\
    s&=&1+(1-j)\left(6+q-\frac{3(1-j)}{1+q}\right)\\
    &~&-3(1+q)\left(1-\frac{4Q^2(1+q)^2\left(5f_{QQQ}+2Qf_{QQQQ}\right)}{3\left(f_Q+2Qf_{QQ}\right)}\right)\label{snap} 
    \end{eqnarray}
In case of the power law $f(Q)$ model using the equations (\ref{decl})-(\ref{snap}) we get the following forms of cosmographic parameters:
\begin{eqnarray}
    q&=&-1-\frac{3}{2}\frac{\Omega_m}{\left(1+mn(2n-1)Q^{n-1}\right)}\\
    j&=&1-\frac{9}{2}\Omega_m^2\frac{mn(n-1)(2n-1)Q^{n-1}}{\left(1+mn(2n-1)Q^{n-1}\right)^3}\\
    s&=&1+(1-j)\left(6+q-\frac{3(1-j)}{1+q}\right)\\
    &~
    &-3(1+q)\left(1-\frac{4Q^{n-1}(1+q)^2mn(n-1)(n-2)(2n-1)}{3\left(1+mn(2n-1)Q^{n-1}\right)}\right)
   \end{eqnarray}
Let the present value of $q,j$ and $s$ be represented by $q_0,j_0$ and $s_0$ respectively. In fig-5 We plot graphs for $q(z),j(z)$ and $s(z)$ for different values of $n>1$.

\begin{figure}[!ht]
 \centering
   \includegraphics[scale=0.74]{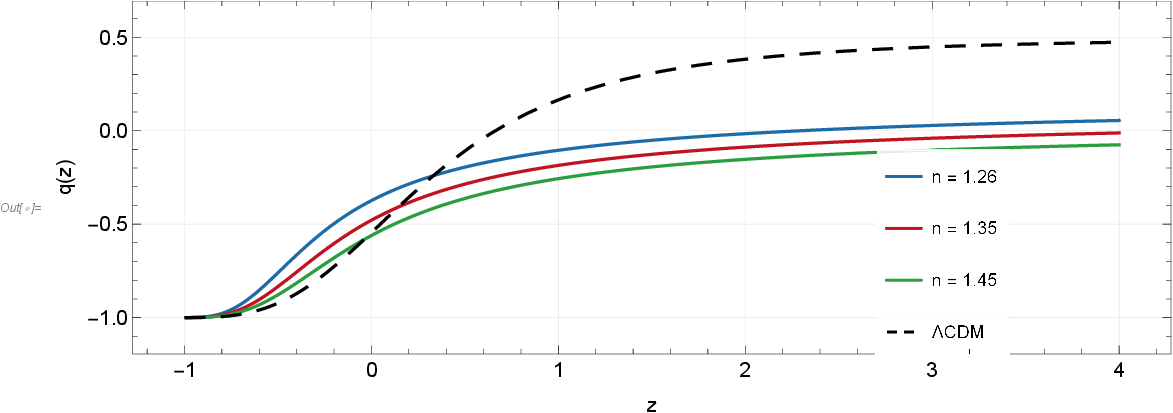}
 \end{figure}
\begin{figure}[!ht]
 \centering
   \includegraphics[scale=0.5]{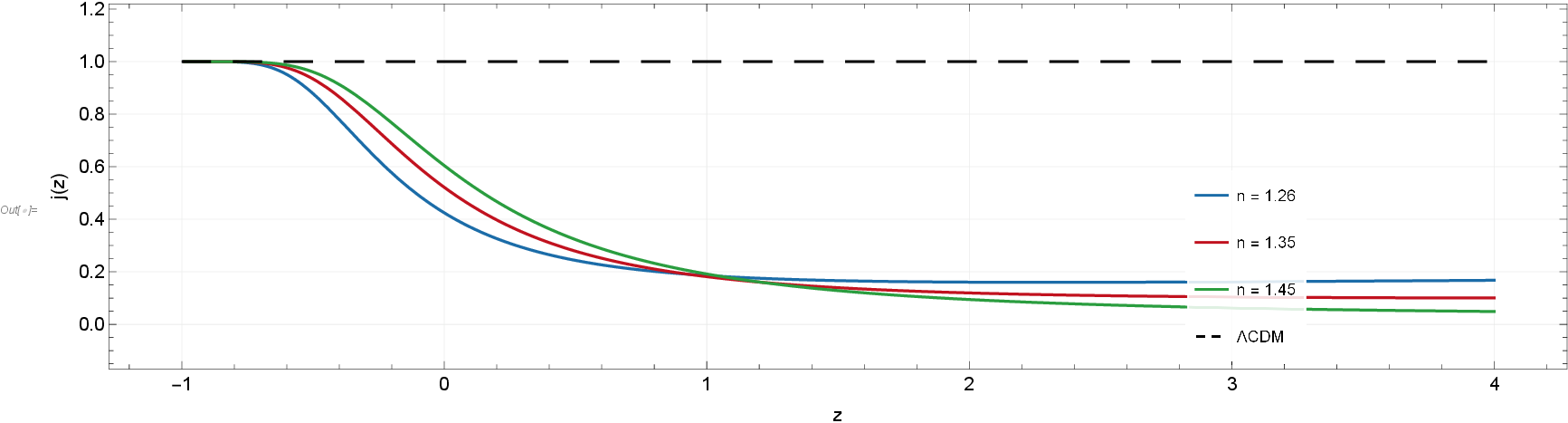}
 \end{figure}
\begin{figure}[!ht]
 \centering
  \includegraphics[scale=0.8]{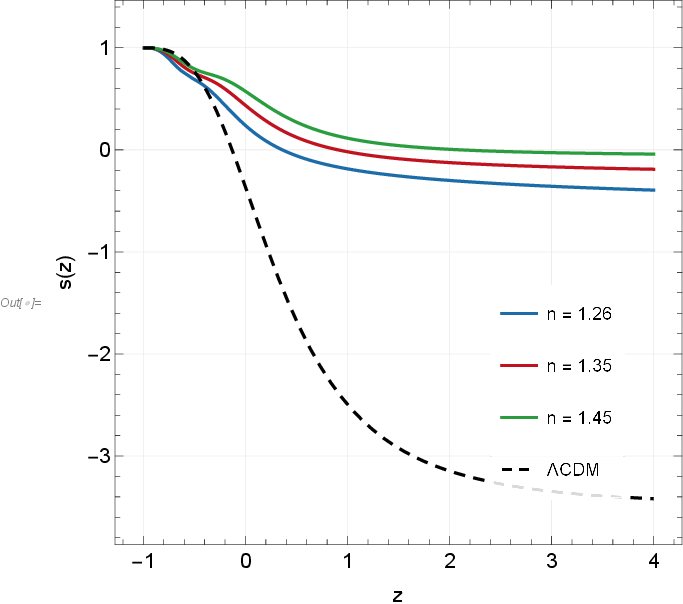}
 \caption{Plots of cosmographic parameter with $H_0=72.26,\Omega_{0m}=0.299$($H(z)+$Pantheon+ data set)}
\end{figure}
 From these graphs of cosmographic parameters we get some interesting findings. We observe that the evolution path of the cosmographic parameters for all values of $n$ are very distinct from that of $\Lambda$CDM model. The deceleration parameter in all cases converges to De Sitter solution ($q=-1$) as $z\rightarrow -1$. Jerk and snap parameters also coincide with $\Lambda$CDM model as $z\rightarrow -1$.   So, for all values of $n$ our $f(Q)$ gravity model and $\Lambda$CDM model are indistinguishable in the far future. But at present and even in near past they quickly separate from each other. In table-\ref{tab:cosmography}, present values of the cosmographic parameters $q,j,s$ and also the transition redshift $z_t$ for each $n$ are shown along with their corresponding values in $\Lambda$CDM model. Here we have considered three specific values of $n$ viz $n=1.26$ which is the best fit value of $n$ derived from observational analysis in section 2, and $n=1.35, n=1.45$. We see that at $n=1.26$, $q_0$ is almost $30\%$ higher and  $j_0$ is almost $60\%$ lower than $\Lambda$CDM model. $q_0,j_0$ gradually get close to $\Lambda$CDM value with the increase of $n$. Infact at $n=1.45$ $q_0$ is very similar to $\Lambda$CDM model, slightly lower than it. Similarly if $n$ is increased, $j_0$ can be made very close to 1. But one thing here is important to mention that, increase in $n$ costs in the value of transition redshift $z_t$. While current observations favours $z_t\approx 0.6 -0.8$  which we see in case of $\Lambda$CDM model ($z_t=0.6632$) , for $n=1.26$ it is 2.3024 i,e even at the observationally best fit value of $n$ the model postpones the onset of acceleration far too early in cosmic history and with the increase of $n$ the value $z_t$ increases sharply. The difference between our $f(Q)$ gravity and $\Lambda$CDM model is more evident in the snap parameter graph. In all the four curves $s=+1$ at $z=-1$, but $\Lambda$CDM  then plunges monotonically to $-3.42$ by $z = 4$, while the $f(Q)$ curves decline only gently and flatten between $-0.04$ and $-0.39$. Across the entire observable range $z > 0$ the snap is unambiguously higher than $\Lambda$CDM for every $n$ considered. 
    
\begin{table}[h]
\centering
\caption{Present value of cosmographic parameters in $f(Q)$ gravity}
\label{tab:cosmography}
\begin{tabular}{|c|c|c|c|c|}
\hline
Model & $q_0$ & $j_0$ & $s_0$ & $z_t$ \\
\hline
$n=1.26$        & $-0.3738$ & $0.4244$ & $0.2391$ & $2.3024 $ \\
\hline
$n=1.35$        & $-0.4789$ & $0.5215$ & $0.4362$ & $4.6071$ \\
\hline
$n=1.45$        & $-0.5608$ & $0.6043$ & $0.5740$ & $15.0648$ \\
\hline
$\Lambda$CDM    & $-0.5455$ & $1$      & $-0.3635$ & $0.6632$ \\
\hline
\end{tabular}
\end{table}

\section{$Om(z)$ Diagnostic}

The $Om(z)$ diagnostic provides an alternative method to test cosmic acceleration assuming an equation of state
\[
p=\rho\omega,
\]
where $\omega$ represents the equation-of-state parameter.

This diagnostic distinguishes different dark energy models such as quintessence, phantom, and $\Lambda$CDM models. Its slope determines the cosmic phase:
\begin{itemize}
\item Positive slope $\Rightarrow$ Phantom region ($\omega<-1$),
\item Negative slope $\Rightarrow$ Quintessence region ($\omega>-1$).
\end{itemize}

The diagnostic is defined as

\begin{equation}
Om(z)=\frac{E^2(z)-1}{(1+z)^3-1},
\end{equation}

where
\[
E(z)=\frac{H(z)}{H_0}.
\]

For the $\Lambda$CDM model, plotting $H^2$ versus $(1+z)^3$ produces a straight line, whereas alternative dark energy models yield curved behaviour. The decreasing nature of $Om(z)$ shown in Fig.~3 indicates a phantom phase at present cosmic time that approaches $\Lambda$CDM behaviour at late times.

\begin{figure}[htbp]
  \centering
  \includegraphics[width=0.7\textwidth]{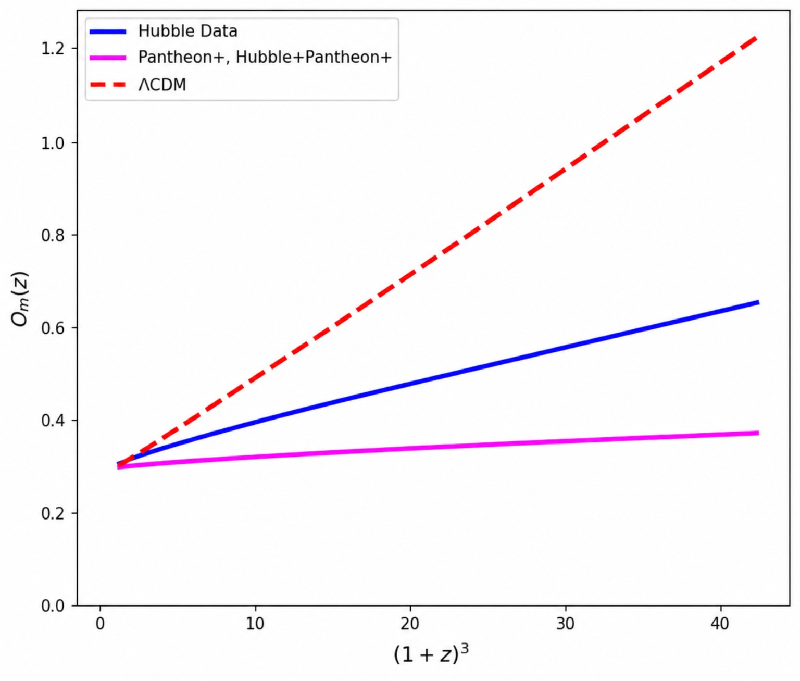}
  \caption{$Om(z)$ diagnostic plot}
  \label{om}
\end{figure}

The $Om(z)$ diagnostic plot in figure-\ref{om} compares the evolution of the cosmological model constrained by different datasets with the standard $\Lambda$CDM scenario. The horizontal axis represents $(1+z)^3$ while the vertical axis shows the $Om(z)$ parameter.The dashed red line corresponds to the standard $\Lambda$CDM model. In the $\Lambda$CDM framework, the $Om(z)$ diagnostic generally behaves as a constant quantity associated with the matter density parameter. The strong linear increase shown here reflects the expected reference evolution used for comparison. The blue curve represents the model constrained using only the Hubble H(z) dataset. This curve increases steadily with redshift but remains below the $\Lambda$CDM prediction throughout the entire redshift range. This indicates that the modified $f(Q)$ model predicts a slower growth of the expansion rate compared with the standard cosmological model when constrained solely by Hubble observations.
The magenta curve corresponds to the combined Pantheon+ and Hubble dataset analysis. This curve evolves much more slowly and stays significantly flatter than both the Hubble-only and $\Lambda$CDM curves. The flatter behavior suggests that the joint dataset favors a cosmological evolution closer to a weakly varying dark-energy dominated universe.The separation between the curves becomes larger at higher redshift values. This implies that deviations from $\Lambda$CDM are more pronounced in the early universe regime. The combined dataset produces the smallest deviation rate, indicating tighter constraints and improved parameter estimation due to the inclusion of supernova observations.

\section{Summary and Conclusion}

In this paper $f(Q)$ model of Universe is examined using current observations and in context of thermodynamics. We considered flat, homogeneous and isotropic universe bounded by apparent horizon. In particular We have investigated the functional form $f(Q)=Q+mQ^n$ where $m,n$ are constants. In our observational analysis we considered three types of data sets viz 31-point Hubble $H(z)$, combined 1701 point Pantheon type SNIa and SH0ESdata and combined Hubble+Pntheon+SHoES data. Using MCMC method we determined the best fit values of model parameters and compared them with that of standard $\Lambda$CDM model. We observed that combined $H(z)+Pantheon+SH0ES$ data set finds $H_0\approx 72.67$ and $\Omega_{0m}\approx 0.229$  for our $f(Q)$ model which are slightly higher than that of the $\Lambda$CDM model $H_0\approx 68.11$ and $\Omega_{0m}\approx 0.323$. For all three types of data sets we have $\chi^2/dof\approx 1$ for our model indicating considerably good fit to current observed data. From error bar plot it comes out that at low redshift $f(Q)$ gravity model almost coincide with $\Lambda$CDM model while at higher redshift their difference is clearly visible. Therefore, at present epoch $f(Q)$ model behaves more similar like $\Lambda$CDM model but deviates significantly in future. The best fit values of the constant parameter $n$ for three data sets lie within $[1.25,1.26]$ respectively. We also plotted a total of 10 contour plots to understand inter-relationship between different model parameters. This fact is reflected in our findings from thermodynamic analysis also. After observational analysis we carried out our investigation for $f(Q)$ gravity from thermodynamic point of view. In  Einstein gravity, the horizon entropy obeys Bekenstein's area law and is equal to $\frac{A}{4G}$ where $A$ is the area of the bounding horizon. In case of modified gravity horizon entropy is not simply the Bekenstein-Hawking entropy but is a generalized form that often includes correction terms. Using Unified law of thermodynamics we found horizon entropy in $f(Q)$ gravity. Then following standard Eckart theory for irreversible thermodynamics we derived first and second time derivative of total entropy $\frac{dS_T}{dt}$ and $\frac{d^2 S}{dt^2}$. We examined validity of generalized second law of thermodynamics(GSLT) and thermodynamic equilibrium for which we need $\frac{dS_T}{dt}\geq 0$ and $\frac{d^2 S}{dt^2}<0$ respectively and accordingly conditions are imposed for general case. Then we applied the results in the power law model of $f(Q)$ gravity : $f(Q)=Q+mQ^n$. We see that at present time, with $H_0=72.26,\Omega_{0m}=0.229$, thermodynamic equilibrium (TE) is satisfied both for positive and negative value of $n$ for all values of thermal conductivity parameter $\lambda\in(0,1)$, but GSLT is valid only when $n\ge 1$., Infact from the region plot-4(c),  we see that within the specific range $1\leq n\leq 2.4$, GSLT is valid but TE fails to satisfy. We prefer this range of $n$ due to three reasons. Firstly, these ranges are calculated keeping parity with current observed values of the cosmographic parameters $H, q$ and $j$. Secondly, within these range strong energy condition (SEC) is violated indicating to current accelerating phase of universe. Lastly due to irreversible nature of the process of heat flow considered in this paper, TE should be violated. We checked our results graphically to show validity of GSLT and violation of TE within the preferred range of model parameters.\\
The behaviour of cosmographic parameters are then checked within the preferred range of $n$. Keeping best fit value of $n=1.26$ as baseline, the cosmographic parameters are studied considering two more values of $n$, viz. $n=1.35$ and $n=1.45$. We see from plots of the cosmographic parameters against redshift $z$, that for every value of $ n$ considered, the deceleration parameter approaches the de Sitter value $q=-1$ as $z\rightarrow -1$ and likewise the jerk and snap parameters approach their $\Lambda$CDM counterparts in the same limit. The model is therefore indistinguishable from $\Lambda$CDM in the asymptotic future, and shares its ultimate fate. The distinction is entirely a feature of the present epoch and the near past. The cosmographic trajectories separate from the $\Lambda$CDM curves rapidly once $z$ becomes positive where the two scenarios are most easily discriminated. At the best-fit $n=1.26$, it is observed that $q_0$ exceeds the $\Lambda$CDM value by roughly $30\%$ and $j_0$ falls short of it by roughly $60\%$. Both quantities move monotonically towards the $\Lambda$CDM values as $n$ is increased. Throughout the entire observable range $z>0$ the snap parameter $s$ in the $f(Q)$ model lies unambiguously above the $\Lambda$CDM prediction for every $n$ considered. Therefore $s$ works as the most promising observational handle for distinguishing the two scenarios.  The overall assesment for this work is that the thermodynamic requirement $n\geq 1$ and the observational best fit $n=1.26$ are mutually consistent i.e the value selected by the data lies comfortably inside the region permitted by the generalised second law of thermodynamics. The only difficulty is the transition redshift. $z_t$ increases with $n$ while the cosmographic agreement also improves with $n$ making the two requirements pull in opposite directions. To overcome this difficulty, in future we shall try some modifications in the  the power law form of $f(Q)$ gravity.

\end{document}